\documentclass[preprint2]{aastex6}
\usepackage{graphicx}
\usepackage{amsmath}	
\usepackage{amssymb} 
\usepackage{float}

\newcommand{\OI}{O\,{\sc i}}

\newcommand{\CaII}{Ca\,{\sc ii}}
\newcommand{\FeII}{Fe\,{\sc ii}}

\begin{document}

\title{Excitation mechanism of \OI~lines in Herbig Ae/Be stars}
\shorttitle{\OI~emission lines in Herbig Ae/Be Stars}

\author{Blesson Mathew\altaffilmark{1}}
\affil{Department of Physics, Christ University \\
  Hosur Road, Bangalore 560029, India}

\affil{Department of Astronomy and Astrophysics\\
Tata Institute of Fundamental Research \\
Homi Bhabha Road, Colaba, Mumbai 400005, India}

\author{P. Manoj}
\affil{Department of Astronomy and Astrophysics\\
Tata Institute of Fundamental Research \\
Homi Bhabha Road, Colaba, Mumbai 400005, India}

\author{Mayank Narang}
\affil{Department of Astronomy and Astrophysics\\
Tata Institute of Fundamental Research \\
Homi Bhabha Road, Colaba, Mumbai 400005, India}

\author{D. P. K. Banerjee}
\affil{Astronomy and Astrophysics Division \\
Physical Research Laboratory \\
Navrangapura, Ahmedabad 380 009, India}

\author{Pratheeksha Nayak}
\affil{Indian Institute of Space Science and Technology (IIST)\\
Trivandrum, India}

\author{S. Muneer}
\affil{Indian Institute of Astrophysics\\
Koramangala, Bangalore 560034, India}

\author{S. Vig}
\affil{Indian Institute of Space Science and Technology\\
Trivandrum, India}

\author{Pramod Kumar S.}
\affil{Indian Institute of Astrophysics\\
Koramangala, Bangalore 560034, India}

\author{Paul K. T.}
\affil{Department of Physics, Christ University \\
  Hosur Road, Bangalore 560029, India}

\and

\author{G. Maheswar}
\affil{Indian Institute of Astrophysics\\
Koramangala, Bangalore 560034, India}

\shortauthors{B. Mathew et al.}
  
\altaffiltext{1}{blesson.mathew@christuniversity.in}

\begin{abstract}
  We have investigated the role of a few prominent excitation mechanisms
  viz. collisional excitation, recombination, continuum fluorescence and
  Lyman beta fluorescence on the \OI~line spectra in Herbig Ae/Be stars.
  The aim is to understand which of them is the central  mechanism that
  explains the observed \OI~line strengths. The study is based on an
  analysis of the observed optical spectra of 62 Herbig Ae/Be stars and near-infrared spectra of
  17 Herbig Ae/Be stars. The strong correlation observed between the line fluxes of \OI~$\lambda$8446
  and \OI~$\lambda$11287, as well as a high positive correlation between the line strengths of
  \OI~$\lambda$8446 and H$\alpha$ suggest that Lyman beta fluorescence is the
  dominant excitation mechanism for the formation of \OI~emission lines in Herbig Ae/Be stars. 
  Further, from an analysis of the emission line fluxes of \OI~$\lambda\lambda$7774, 8446,
  and comparing the line ratios with those predicted by theoretical models,
  we assessed the contribution of collisional
  excitation in the formation of \OI~emission lines. 
\end{abstract}


\section{Introduction } 
\label{sec intro}

Herbig Ae/Be (HAeBe) stars are intermediate mass ($2~M_\odot \leq M \leq 8~M_\odot$)
pre-main sequence stars with
accretion disks, the innermost regions of which also act as a
reservoir for the production of major emission lines seen in the
optical and infrared spectra \citep{Herbig60,Hillenbrand92,Waters98}.
HAeBe stars were first discussed as a distinct group of objects
by \citet{Herbig60}, who noted that they were stars of spectral 
type A or B with emission lines, located in an obscured region 
and often accompanied by a surrounding nebulosity. The present
working definition of HAeBe stars includes,
(a) pre-main sequence stars of A$-$F spectral type, displaying emission lines 
in their spectra and (b) show a significant IR excess due to 
hot or cool circumstellar dust shell or a combination of 
both \citep{The94,Waters98,Vieira03}. There have been extensive
spectroscopic studies of HAeBe stars in the literature
\citep[e.g.][]{Hamann92,Hernandez04,Manoj06}; particularly important are the recent
studies by the X-Shooter team \citep{Mendigutia11,Mendigutia12,Fairlamb15,Fairlamb17}.
Most of these studies have been devoted to H$\alpha$ line analysis, the
most prominent emission feature seen in the spectra of HAeBe
stars \citep{Finkenzeller84,Hamann92}. In the present study, 
we focus on the \OI~emission lines
in the optical and near-infrared (1$-$2.5 micron) spectra in HAeBe stars.

\OI~$\lambda$8446 is the most prominent \OI~emission
line seen in the optical spectrum of HAeBe stars. This emission line results from the
3$s^3S^0$~$-$3p$^3P$ transition and is seen as a triplet at high resolution,
with wavelength values of 8446.25, 8446.36 and 8446.76 \AA. It is present in
the spectra of a wide variety of astrophysical sources such as planetary nebulae,
novae and Seyfert galaxies. A number of studies have addressed  the question of
excitation mechanisms of \OI~emission lines  in various astrophysical objects.
Prominent mechanisms discussed for the formation of \OI~lines are  collisional
excitation, recombination, continuum fluorescence and Lyman beta (Ly$\beta$) fluorescence.
For example, \citet{Grandi75b} showed that starlight
continuum fluorescence is the favored excitation mechanism for the 
\OI~line in the Orion nebula whereas in Seyfert 1 galaxies it is excited
by Lyman $\beta$ fluorescence \citep{Grandi80}. In novae, \citet{Strittmatter77} identify 
Lyman $\beta$ fluorescence as the dominant excitation mechanism; a conclusion that has been
supported by studies of several other novae \citep[e.g][and references therein]{Ashok06,Banerjee12}.
Ly$\beta$ fluorescence is identified as the dominant contributor to the
emission strength of \OI~$\lambda$8446 line in classical Be (hereafter CBe) stars, whether it is
isolated \citep{Slettebak51,Mathew12b} or part of an X-ray binary
system \citep{Mathew12a}.
\citet{Bhatia95} and \citet{Kastner95} provided a theoretical framework of \OI~excitation and derived
the expected line ratios of the prominent \OI~lines, when collisional excitation and
Ly$\beta$ fluorescence \citep[referred as photoexcitation by accidental resonance (PAR process) in][]{Bhatia95}
are the dominant excitation mechanisms. From a comparative analysis of the theoretical estimates with the
observed emission strengths of \OI~$\lambda\lambda$ 7774, 8446, 11287 and 13165,
\citet{Mathew12b} demonstrated that Ly$\beta$ fluorescence  is the
dominant excitation mechanism for the production of
\OI~$\lambda\lambda$ 8446, 11287 lines in CBe stars. CBe stars share similar spectral
characteristics with HAeBe stars, such as emission lines of
H$\alpha$, \OI, \FeII~and \CaII~triplet. It is worth exploring whether both CBe and HAeBe stars
share similar excitation mechanism for the formation of \OI~lines. There could be considerable difference
between the \OI~line forming regions in both the stellar systems. CBe stars are found to have a
circumstellar gaseous decretion disk wherein \OI~$\lambda$8446 line is formed at a mean radial distance
of $\sim$8 $R_{\star}$, considering Keplerian motion \citep{Mathew12b}. However, the location of the origin of
\OI~$\lambda$8446 line in HAeBe is far from clear. Most of the accretion
related emission lines in HAeBe stars (e.g. H$\alpha$, Pa$\beta$, Br$\gamma$) are thought to be formed in the
magnetospheric accretion columns \citep{Muzerolle04}. This work is an attempt to bring more clarity to our
understanding of the formation mechanisms of \OI~emission lines in HAeBe stars. 
  
The paper is organized as follows. In Section 2 we present the optical and near-infrared (near-IR)
spectroscopic observations
carried out over a period of 3 years and describe the data reduction techniques employed.
We describe the methods and the python routines used for the spectral analysis and
to estimate line flux in Section 3. The dominant excitation mechanism
for the formation of \OI~lines in HAeBe stars is
evaluated in Section 4. The main results of the paper are summarized in Section 5. 

\section{Observations and data reduction}

The optical spectroscopic observations were carried out using the Himalayan Faint Object Spectrograph Camera (HFOSC)
mounted on the 2-m Himalayan Chandra Telescope (HCT)\footnote{http://www.iiap.res.in/iao/hfosc.html}.
The spectroscopic observations were obtained with Grism 8 in combination with 167$l$ slit
(1.92$\arcsec$ wide and 11$\arcmin$ long), providing an effective resolving power of $\sim$ 1050.
The spectral coverage is from 5500 to 9000 \AA, which included the spectral lines relevant to this
study, viz., H$\alpha$, \OI~$\lambda$7774 and \OI~$\lambda$8446. After each on-source exposure,
FeNe lamp spectra were obtained for wavelength calibration. We have followed the regular procedure of
reducing the spectra after bias subtraction and flat-field correction using the standard tasks in
Image Reduction and Analysis Facility (IRAF)\footnote{IRAF is distributed by the National Optical
  Astronomy Observatories, which are operated by the Association of Universities for Research in
  Astronomy, Inc., under cooperative agreement with the National Science Foundation}.

\begin{figure*}[ht]
\begin{center}
\figurenum{1}
\includegraphics[width=18cm]{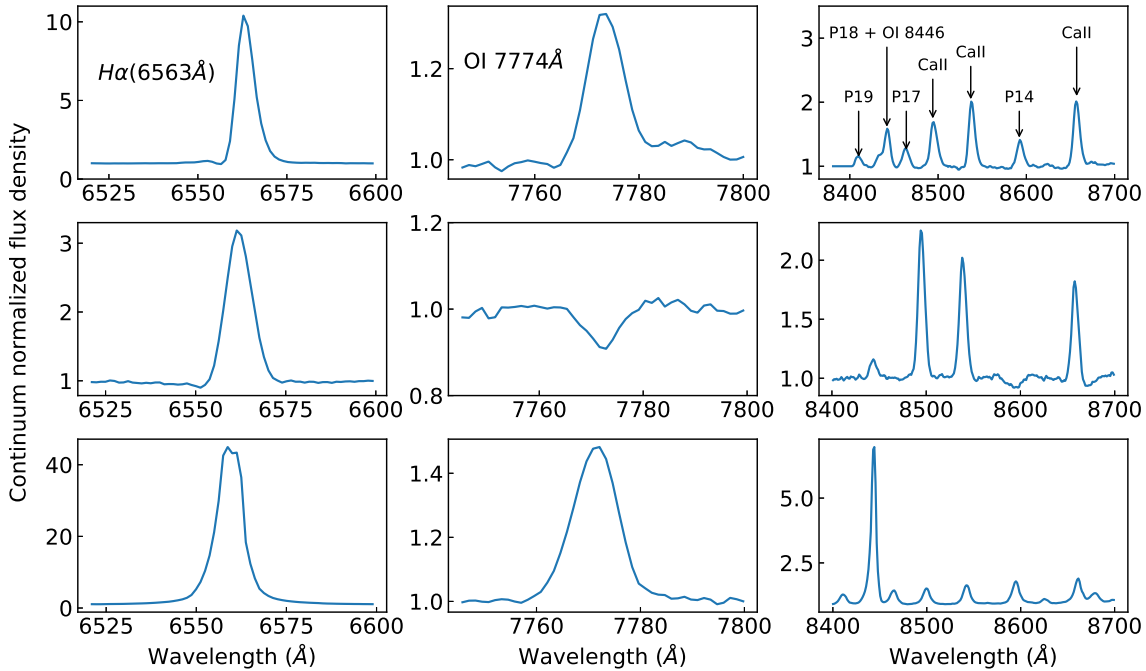}
\caption{Observed spectra of V594 Cas, LkH$\alpha$ 233 and MWC 297 (top to bottom).
  The line profiles of H$\alpha$, \OI~$\lambda$7774 and \OI~$\lambda$8446 are shown in
  each case (left to right). \CaII~triplet (8498, 8542, 8662 \AA) lines are seen in most
  spectra and appears to be blended with Paschen lines when they are in emission.} 
\label{figure1}
\end{center}
\end{figure*}

Near-IR spectra were obtained using the TIFR Near Infrared Spectrometer and Imager (TIRSPEC),
mounted on the HCT. The spectra were obtained in $Y$ and $J$ passbands, at a resolving power
of 1200. The observations were performed in the dithered mode. Argon lamp spectra taken after
each on-source exposure is used for wavelength calibration. An appropriate telluric standard
(of early A spectral type) is observed at nearby airmass to the target object. The spectra of
the target and the standard are reduced in a standard manner with the tasks in IRAF. For
telluric correction, we removed the hydrogen lines from the telluric standard spectrum, which
is then used to divide the object spectrum. The resultant object spectrum is multiplied with
the blackbody corresponding to the spectral type of the telluric standard in order to preserve
the continuum of the target spectrum. The log of optical and infrared spectroscopic
observations is given in Table \ref{table1}. 

The sample of HAeBe stars observed were drawn from a larger list of 142 HAeBe stars
that we compiled from literature \citep{The94,Manoj06,Fairlamb15}.
Given the location of the observatory and the limiting magnitude of the
spectrograph-telescope combination, we were able to obtain the optical spectra
of 56 HAeBe stars and near-IR spectra of 19 HAeBe stars.
The observations were carried out over a period of 3 years, from 2014 to 2017.
To increase the sample size of the present study we have included the optical spectra of HAeBe stars
from \citet{Manoj06}, which were observed with similar observation setup.
Thus we have optical spectra for a total of 62 HAeBe stars and near-IR spectra for 19 HAeBe stars.
As a representative sample, we show H$\alpha$, \OI~$\lambda$ $\lambda$7774, 8446
line profiles of V594 Cas, LkH$\alpha$ 233 and MWC 297 in Figure \ref{figure1}.

The $B$, $V$, $R_C$ magnitudes, total extinction ($A_V$),
  spectral type and effective temperature ($T_\mathrm{eff}$) of
  62 HAeBe stars are listed in Table \ref{table2}. The spectral type is
  converted to $T_\mathrm{eff}$ using the tabulated information
  in \citet{Pecaut13}. We compiled the photometric data
  from various sources in literature, whose references are given in Table \ref{table2}.
  For some of the sources $R$ magnitudes are in Johnson system, which are
  converted to the Cousins system following \citet{Bessell83}.
  The color excess, $E(B-V)$, is calculated from the observed $(B-V)$ colors and the
  intrinsic colors corresponding to each spectral type, from the table listed
  in \citet{Pecaut13}. Further, we
  calculated $A_V$ from $E(B-V)$ considering a total-to-selective extinction value,
  $R_V$ = 5. It has been demonstrated from various studies \citep[c.f.][]{Hernandez04}
  that $R_V$ = 5 is the preferred value in the analysis of HAeBe stars,
  suggesting grain growth in the disk of HAeBe stars \citep[e.g.][]{Gorti93,Manoj06}).  

\section{Analysis}

\subsection{Classification based on \OI~line profiles}

From the observed spectra, we find that \OI~lines, both in optical and infrared, are seen in emission as well
as in absorption. We adopted the classification scheme proposed 
by \citet{Felenbok88}, wherein Group I stars have both 
\OI~$\lambda$8446 and \OI~$\lambda$7774 in emission; Group II sources are those with both lines
in absorption; Group III is the case when \OI~$\lambda$8446 is in emission and \OI~$\lambda$7774
in absorption. We found 23 stars belonging to Group I, 16 in Group II and 23 in Group III classes.
Similar classification scheme is applied to the infrared spectra. Although \OI~emission is evident
among Group I stars, after subtracting the photospheric component, a net emission is seen in some of
the Group II and Group III stars. For the current sample, we found net emission
in \OI~$\lambda \lambda$7774, 8446 for 31 and 54 stars, respectively whereas 17 sources
show net emission in \OI~$\lambda \lambda$11287 and 13165. 

\begin{deluxetable*}{lccccc}
\tablecaption{Log of spectroscopic observations \label{table1}}
\tablehead{
\colhead{Object} & \colhead{Date of } & \colhead{Optical} & \colhead{Date of } & \colhead{$Y$ band}     & \colhead{$J$ band} \\
\colhead{}       & \colhead{Optical Observations}     & \colhead{Exp.time (s)} & \colhead{IR Observations}     & \colhead{Exp.time (s)} & \colhead{Exp.time (s)}}
\colnumbers
\startdata
51 Oph     & 2014 May. 19 & 60 & -- & --& -- \\
AB Aur     & 2017 Jan. 21 & 40 & 2013 Dec. 11 & 600 & 600\\ 
AS 442     & 2016 Aug. 10 & 300 & -- & -- & -- \\ 
AS 443     & 2016 Aug. 10 & 420 & -- & -- & -- \\  
AS 505     & 2016 Aug. 10 & 300 & 2016 Nov. 21 & 600 & 600\\  
BD+30 549  & 2016 Nov. 22 & 400 & -- & -- & -- \\   
BD+40 4124 & 2014 Jun. 19 & 300 & -- & -- & -- \\ 
BD+65 1637 & 2016 Aug. 10 & 300 & -- & -- & -- \\   
CQ Tau     & 2016 Nov. 23 & 300 & 2016 Nov. 21 & 600 & 600\\  
HBC 334    & 2017 Jan. 03 & 1800 & -- & -- & -- \\   
HBC 551    & 2014 Feb. 25 & 1200 & -- & -- & -- \\   
HD 141569  & 2014 Jun. 19 & 300 & -- & -- & -- \\  
HD 142666  & 2014 Feb. 24 & 300 & -- & -- & -- \\
           & 2014 May. 19 & 300 & -- & -- & -- \\  
HD 144432  & 2014 May. 19 & 300 & -- & -- & -- \\  
HD 145718  & 2016 May. 15 & 300 & -- & -- & -- \\  
HD 150193  & 2014 May. 19 & 600 & -- & -- & -- \\
           & 2014 Jun. 19 & 300 & -- & -- & -- \\ 
HD 163296  & 2016 May. 15 & 30 & 2016 May. 15 & 120 & 120 \\ 
HD 169142  & 2014 May. 19 & 300 & -- & -- & -- \\ 
HD 190073  & 2014 Oct. 02 & 60  & -- & -- & -- \\ 
HD 200775  & 2017 Jan. 22 & 30 & 2016 Nov. 20 & 120 & 120  \\ 
           & -- & -- & 2017 Jan. 22 & 320 & 320 \\
HD 216629  & 2016 Aug. 10 & 30 & -- & -- & -- \\ 
HD 245185  & 2017 Jan. 03 & 600 & -- & -- & -- \\  
HD 250550  & 2017 Jan. 21 & 600  & 2017 Jan. 22 & 600 & 600 \\  
HD 259431  & 2015 Dec. 16 & 60 & 2016 Nov. 21 & 600 & 480  \\ 
HD 31648   & 2016 Nov. 20 & 60 & 2016 Nov. 20 & 180 & 240\\  
HD 35187   & 2017 Jan. 21 & 60 & -- & -- & -- \\  
HD 35929   & 2017 Jan. 21 & 120 & 2016 Nov. 21 & 480 & 480\\ 
HD 36112   & 2014 Feb. 24 & 180 & 2014 Feb. 24 & 120 & 120\\
           & 2015 Jan. 27 & 90 & 2016 Nov. 21 & 360 & 360 \\ 
HD 37490  & 2017 Jan. 21 & 20  & 2017 Jan. 21 & 320 & 400 \\
HD 37806  & 2016 Nov. 22 & 30  & -- & -- & -- \\
HD 52721  & 2016 Nov. 22 & 30  & 2016 Nov. 21 & 180 & 180\\
HD 53367  & 2016 Nov. 23 & 90  & 2016 Nov. 21 & 180 & 180  \\
HK Ori    & 2017 Jan. 03 & 600  & -- & -- & -- \\
LkHa 167  & 2016 Sep. 25 & 1200 & -- & -- & -- \\
LkHa 198  & 2015 Dec. 16 & 1200 & -- & -- & -- \\
LkHa 224  & 2016 May. 16 & 900  & -- & -- & -- \\
LkHa 233  & 2014 Oct. 02 & 1800 & -- & -- & -- \\
LkHa 234  & 2016 Aug. 10 & 600  & -- & -- & -- \\
LkHa 257  & 2016 Aug. 10 & 900  & -- & -- & -- \\
MWC 1080  & 2014 Aug. 17 & 180   & 2014 Aug. 17 & 200 & 200  \\
          & 2014 Nov. 02 & 180  & -- & -- & -- \\
MWC 297   & 2014 May. 19 & 360  & 2014 Jun. 19 & 120 & 80  \\
          & 2014 Jun. 19 & 600  & 2014 Aug. 17 & 200 & 120 \\
PDS 174   & 2017 Jan. 03 & 900  & -- & -- & -- \\
PX Vul    & 2014 Oct. 01 & 600  & -- & -- & -- \\
SV Cep    & 2016 Aug. 10 & 300  & -- & -- & -- \\
UX Ori    & 2017 Jan. 03 & 60   & -- & -- & -- \\
UY Ori    & 2017 Jan. 03 & 900  & -- & -- & -- \\
V1012 Ori & 2017 Jan. 03 & 900  & -- & -- & -- \\
V1366 Ori & 2017 Jan. 03 & 600  & -- & -- & -- \\
V376 Cas  & 2014 Oct. 02 & 1800 & -- & -- & -- \\
V594 Cas  & 2014 Oct. 01 & 120  & 2016 Nov. 20 & 300 & 400 \\
          & 2014 Nov. 03 & 180  & -- & -- & -- \\
V699 Mon  & 2016 Nov. 23 & 600  & -- & -- & -- \\
VV Ser    & 2016 May. 15 & 1200 & -- & -- & -- \\
VY Mon    & 2017 Jan. 22 & 900  & 2017 Jan. 22 & 500 & 600\\
WW Vul    & 2016 Aug. 10 & 600  & -- & -- & -- \\
          & 2016 May. 15 & 600  & -- & -- & -- \\
XY Per    & 2016 Nov. 22 & 300  & 2017 Jan. 22 & 400 & 600  \\
Z CMa     & 2014 Feb. 24 & 180  & 2014 Feb. 24 & 120 & 40 \\
\enddata
\end{deluxetable*}

\subsection{Flux measurement of \OI~and H$\alpha$ emission lines}

In this section we describe the method we used to measure the line fluxes of H$\alpha$,
\OI~$\lambda\lambda$7774, 8446, 11287 and 13165 lines from the wavelength
calibrated optical and near-IR spectra. The procedure can be summarized as,
(i) estimating the equivalent width of the lines of interest from a Gaussian profile fit
using LMFIT routine in Python, (ii) removing the contribution of Paschen P18 line from \OI~$\lambda$8446,
(iii) accounting for photospheric absorption using synthetic spectra,
(iv) estimation of continuum flux at the wavelength region corresponding to H$\alpha$ and \OI~lines,
and (v) the calculation of extinction corrected line flux from the equivalent width and the continuum flux. 

\subsubsection{Estimation of line equivalent width}

We estimated the equivalent width (EW) of \OI~$\lambda\lambda$7774, 8446, 11287, 13165 and H$\alpha$ lines
using the LMFIT module on the continuum subtracted, continuum
normalized spectra. LMFIT, which is based on an Marquardt Levenberg 
nonlinear least squares minimization algorithm, was used to fit gaussians to the profiles. 

\subsubsection{Removal of Paschen line (P18) contribution from \OI~$\lambda$8446}
  
For the spectral resolution of our observations, the line profiles of \OI~$\lambda$8446
and Paschen 18 (P18; 8437 \AA) are
blended (see Figure \ref{figure1}). We proposed a method in \citet{Mathew12b} to
deblend the P18 contribution from the net
EW in the study of CBe stars, which will be employed here as well.
The Paschen line strengths show a monotonic increase with wavelength and then display a trend of
flattening out around P17 and beyond \citep{Briot81}. Hence it is reasonable to obtain the EW of P18 by
linearly interpolating between the measured EW of P17 (8467 \AA) and P19 (8413 \AA) \citep[see][]{Mathew12b}.
This value is subtracted from the combined EW of \OI~$\lambda$8446 and P18 to obtain
the intrinsic EW of \OI~$\lambda$8446.

\subsubsection{Accounting for Photospheric absorption}

The equivalent widths calculated from emission lines
needs to be corrected for the photospheric absorption. The strength of the absorption component
is estimated from the synthetic spectrum corresponding to the spectral type of the central star
from \citet{Munari05}, which are calculated
from the SYNTHE code \citep{Kurucz93}, using NOVER models as the input stellar
atmospheres \citep{Castelli97}. The EW of underlying absorption component for H$\alpha$,
\OI~$\lambda$7774 and \OI~$\lambda$8446 is estimated using the synthetic spectra
corresponding to the spectral type of the star. Since the synthetic spectra of \citet{Munari05} do not cover the
infrared spectral region, we used NextGen (AGSS2009) theoretical spectra \citep{Hauschildt99} for the analysis of
\OI~$\lambda\lambda$11287, 13165 line profiles. The equivalent width of the photospheric
absorption is subtracted from the EW of the observed emission line to obtain the net equivalent width. 

\subsubsection{Estimation of line fluxes}

The equivalent width of \OI~emission lines, corrected for photospheric absorption, needs to be multiplied
with the underlying stellar continuum flux density to obtain the line flux.
We are taking the extinction corrected $R$-band flux density as a proxy for the
continuum flux density underlying H$\alpha$ line. The method of calculating the
continuum flux density at \OI~emission lines from H$\alpha$ line is described below. 
The continuum flux density at H$\alpha$ is given as, 
$$F_{\nu,cont}(H\alpha) = F_{\nu,0} \times 10^{(\frac{-R_0}{2.5})}$$ where
  $F_{\nu,0} = 3.08 \times 10^{-23}~W\ m^{-2}\ Hz^{-1}$ and R$_0$ is the extinction corrected
  $R_C$ magnitude. The extinction in $R$-band, $A_R$, is estimated from
  A$_V$ using the extinction curve of \citet{McClure09}.

\begin{deluxetable*}{llccccccc}
\tablecaption{List of compiled stellar parameters for analysis of optical lines \label{table2}}
\tablehead{
  \colhead{Source} &  \colhead{Sp. type}  &  \colhead{Ref. Sp. type}  & \colhead{$T_{eff}$ (K)} &   \colhead{$V$} &  \colhead{$B-V$} &  \colhead{$R_C$}  &\colhead{Ref. Photometry} &\colhead{$A_V$}   
}
\colnumbers
\startdata
51 Oph     & B9.5 IIIe     & 1   & 10400 & 4.78  & 0.03  & 4.75  & 1   & 0.4   \\
AB Aur     & A1            & 1   & 9200  & 7.05  & 0.12  & 6.92  & 1   & 0.39  \\
AS 442     & B8Ve          & 14  & 12500 & 10.9  & 0.66  & 10.18 & 3   & 3.85  \\
AS 443     & B2            & 1   & 20600 & 11.35 & 0.66  & 10.78 & 1   & 4.35  \\
AS 505     & B5Vep         & 15  & 15700 & 10.85 & 0.43  & 10.66 & 4   & 2.93  \\
BD+30 549  & B8p           & 16  & 12500 & 10.56 & 0.35  & 10.42 & 4   & 2.3   \\
BD+40 4124 & B3            & 1   & 17000 & 10.69 & 0.78  & 9.92  & 1   & 4.79  \\
BD+46 3471 & A0            & 1   & 9700  & 10.13 & 0.4   & 9.8   & 1   & 2     \\
BD+65 1637 & B4            & 1   & 16700 & 10.18 & 0.39  & 9.79  & 1   & 2.78  \\
BO Cep     & F4            & 1   & 6640  & 11.6  & 0.56  & 11.21 & 1   & 0.74  \\
CQ Tau     & F3            & 1   & 6720  & 10.26 & 0.79  & 9.72  & 1   & 2.01  \\
HBC 334    & B3            & 1   & 17000 & 14.52 & 0.57  & 13.95 & 1   & 3.74  \\
HBC 551    & B8            & 1   & 12500 & 11.81 & 0.26  & 11.54 & 1   & 1.85  \\
HD 141569  & A0Ve          & 1   & 9700  & 7.1   & 0.1   & 7.03  & 1   & 0.5   \\
HD 142666  & A8Ve          & 1   & 7500  & 8.67  & 0.5   & 8.34  & 1   & 1.25  \\
HD 144432  & A9IVe         & 1   & 7440  & 8.17  & 0.36  & 7.92  & 1   & 0.53  \\
HD 145718  & A5Ve          & 8   & 8080  & 9.1   & 0.52  & 8.79  & 2   & 1.8   \\
HD 150193  & A2IVe         & 1   & 8840  & 8.64  & 0.49  & 8.28  & 1   & 2.08  \\
HD 163296  & A1Vep         & 1   & 9200  & 6.88  & 0.09  & 6.82  & 1   & 0.24  \\
HD 169142  & A5Ve          & 1   & 8080  & 8.15  & 0.28  & 7.95  & 1   & 0.6   \\
HD 179218  & A0IVe         & 1   & 9700  & 7.39  & 0.08  & 7.33  & 1   & 0.4   \\
HD 190073  & A2IVe         & 1   & 8840  & 7.73  & 0.13  & 7.7   & 1   & 0.28  \\
HD 200775  & B3            & 1   & 17000 & 7.37  & 0.41  & 7.01  & 1   & 2.94  \\
HD 216629  & B3IVe+A3      & 17  & 17000 & 9.32  & 0.45  & 9.11  & 4   & 3.14  \\
HD 245185  & A1            & 1   & 9200  & 9.94  & 0.1   & 9.88  & 1   & 0.29  \\
HD 250550  & B9            & 1   & 10700 & 9.54  & 0.07  & 9.41  & 1   & 0.7   \\
HD 259431  & B6            & 1   & 14500 & 8.73  & 0.27  & 8.36  & 1   & 2.05  \\
HD 31648   & A3Ve          & 1   & 8550  & 7.7   & 0.2   & 7.59  & 1   & 0.55  \\
HD 35187   & A2e+A7        & 1   & 8840  & 8.17  & 0.22  & 76.4  & 1   & 0.73  \\
HD 35929   & F2III         & 1   & 6810  & 8.13  & 0.42  & 7.87  & 1   & 0.23  \\
HD 36112   & A5IVe         & 1   & 8080  & 8.34  & 0.26  & 8.16  & 1   & 0.5   \\
HD 37490   & B2            & 5   & 20600 & 4.57  & -0.11 & 4.59  & 5   & 0.5   \\
HD 37806   & A2Vpe         & 1   & 8840  & 7.95  & 0.04  & 7.89  & 1   & -0.17 \\
HD 38120   & B9            & 1   & 10700 & 9.01  & 0.06  & 8.93  & 1   & 0.65  \\
HD 52721   & B1            & 5   & 26000 & 6.62  & 0.06  & 6.53  & 5   & 1.69  \\
HD 53367   & B0IV/Ve       & 9   & 31500 & 6.95  & 0.42  & 6.67  & 2   & 3.64  \\
HK Ori     & A4+G1V        & 1   & 8270  & 11.71 & 0.56  & 11.2  & 1   & 2.1   \\
IP Per     & A6            & 1   & 8000  & 10.47 & 0.33  & 10.24 & 1   & 0.8   \\
LkHa 167   & A2            & 6   & 8840  & 15.06 & 1.42  & 14.32 & 4   & 6.73  \\
LkHa 198   & B9            & 1   & 10700 & 14.18 & 0.95  & 13.31 & 1   & 5.1   \\
LkHa 224   & F9            & 1   & 6040  & 14.07 & 1.44  & 12.98 & 1   & 4.44  \\
LkHa 233   & A4            & 1   & 8270  & 13.56 & 0.84  & 12.91 & 1   & 3.5   \\
LkHa 234   & B7            & 1   & 14000 & 12.21 & 0.9   & 11.49 & 1   & 5.14  \\
LkHa 257   & B5            & 7   & 15700 & 13    & 0.3   & 12.72 & 4   & 2.28  \\
MWC 1080   & B0eq          & 1   & 31500 & 11.52 & 1.34  & 10.39 & 1   & 8.24  \\
MWC 297    & B1.5Ve        & 10  & 24800 & 12.03 & 2.24  & 10.18 & 2   & 12.46 \\
PDS 174    & B3e           & 11  & 17000 & 12.84 & 0.81  & 12.18 & 2   & 4.94  \\
PX Vul     & F3            & 1   & 6720  & 11.54 & 0.83  & 11.12 & 1   & 2.21  \\
R Cra      & A0            & 1   & 9700  & 12.2  & 1.09  & 11.03 & 1   & 5.45  \\
SV Cep     & A0            & 1   & 9700  & 10.98 & 0.39  & 10.68 & 1   & 1.95  \\
UX Ori     & A3            & 1   & 8550  & 10.4  & 0.33  & 10.13 & 1   & 1.2   \\
UY Ori     & B9            & 12  & 10700 & 12.79 & 0.37  & 12.56 & 2   & 2.2   \\
V1012 Ori  & A3e           & 13  & 8550  & 12.04 & 0.42  & 11.61 & 2   & 1.65  \\
V1366 Ori  & A0            & 1   & 9700  & 9.89  & 0.16  & 9.8   & 1   & 0.8   \\
V376 Cas   & B5e           & 1   & 15700 & 15.55 & 1.13  & 14.59 & 1   & 6.43  \\
V594 Cas   & B8            & 1   & 12500 & 10.58 & 0.56  & 10.03 & 1   & 3.35  \\
V699 Mon   & B6            & 1   & 14500 & 10.54 & 0.54  & 10.06 & 1   & 3.4   \\
VV Ser     & B6            & 1   & 14500 & 11.92 & 0.93  & 11.11 & 1   & 5.35  \\
VY Mon     & B8            & 1   & 12500 & 13.47 & 1.55  & 12.19 & 1   & 8.3   \\
WW Vul     & A3            & 1   & 8550  & 10.74 & 0.44  & 10.45 & 1   & 1.75  \\
XY Per     & A5            & 1   & 8080  & 9.21  & 0.49  & 8.86  & 1   & 1.65  \\
Z CMa      & B0 IIIe       & 1   & 31500 & 9.47  & 1.27  & 8.63  & 1   & 7.89 \\
\enddata
\vspace{0.2cm}
References. (1) \citet{Manoj06}; (2) \citet{Fairlamb15}; (3) \citet{Mendigutia12} ;
(4) \citet{Zacharias04}; (5)\citet{Hillenbrand92}; 
(6) \citet{Cohen79}; (7) \citet{Liu11}; 
(8) \citet{Carmona10}; \\(9)\citet{Tjin01};  (10) \citet{Drew97}; (11) \citet{Gandolfi08}; (12) \citet{Vieira03};\\
(13) \citet{Lee07}; (14) \citet{Mora01}; (15) \citet{Garrison70}; (16) \citet{McDonald17}; (17) \citet{Skiff14};

\end{deluxetable*}

The continuum flux densities of \OI~lines 7774 and 8446 are estimated from
H$\alpha$ continuum flux using the relation,\\
$$F_{\lambda, cont}(7774)= \Bigg(\frac{F_{\lambda c}(7774)}{F_{\lambda c}(H\alpha)}\Bigg) \times F_{\lambda,cont}(H\alpha)$$\\
$$F_{\lambda, cont}(8446)= \Bigg(\frac{F_{\lambda c}(8446)}{F_{\lambda c}(H\alpha)}\Bigg) \times F_{\lambda,cont}(H\alpha)$$\\
The ratio of continuum flux densities, $\frac{F_{\lambda c}(7774)}{F_{\lambda c}(H\alpha)}$
and $\frac{F_{\lambda c}(8446)}{F_{\lambda c}(H\alpha)}$ are calculated using the synthetic spectra given in
\citet{Munari05}. The line fluxes of \OI~$\lambda$7774 and \OI~$\lambda$8446 are
obtained by taking a product of the continuum flux density with the measured equivalent width. 
Similarly, the continuum flux density in the near-IR region is
calculated from the extinction corrected $J$ magnitudes of HAeBe stars.
Further, the flux values of \OI~$\lambda$11287 \& \OI~$\lambda$13165 are
calculated from the continuum flux densities and the measured equivalent widths. 

\section{Results \& Discussion}

\subsection{Excitation mechanisms for \OI~emission}

The excitation mechanisms contributing to \OI~emission that are discussed
extensively in literature are recombination, collisional excitation,
continuum fluorescence and Lyman $\beta$ fluorescence \citep{Grandi75b,Strittmatter77,Grandi80,Ashok06,Banerjee12}.
In this section, we assess which one of the above is the dominant mechanism for the production
of \OI~lines in HAeBe stars. 

\subsubsection{Recombination}

One of the possible formation mechanism of permitted \OI~emission lines is through
recombination followed by cascade from higher ionization states.
However, recombination process alone is not sufficient
to explain the strength of \OI~lines in systems such as Orion nebula \citep{Grandi75b}. 

\begin{figure}
\figurenum{2}  
\includegraphics[width=8cm,height=8cm]{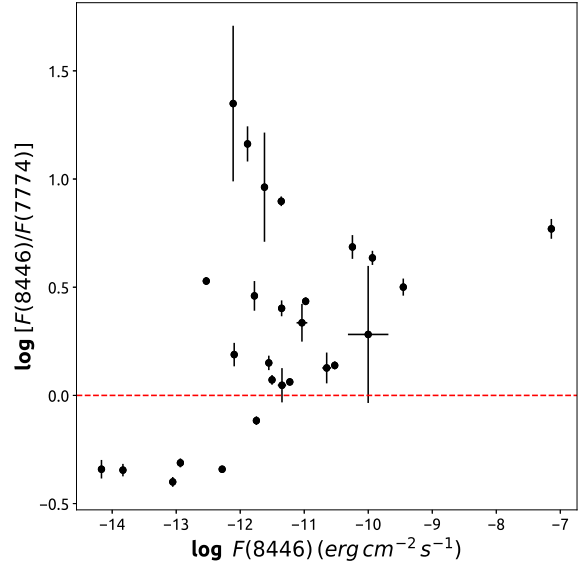}
\caption{Log-Log plot of F($\lambda$8446)/F($\lambda$7774) vs  F($\lambda$8446):
  The error bars are indicated. In 77\% cases, F($\lambda$8446) $>$ F($\lambda$7774).}
\label{figure2}
\end{figure}

If the recombination process is the dominant mechanism, then
the emission strengths of $\lambda$7774 and
$\lambda$8446 should follow the ratio of statistical weights,
i.e., F($\lambda$7774)/F($\lambda$8446) = 5/3 \citep{Strittmatter77,Grandi80}.
So, if recombination operates in HAeBe stars, we should
expect \OI~$\lambda$7774 to be stronger than \OI~$\lambda$8446.
The flux ratio of \OI~$\lambda$7774 and $\lambda$8446 is shown as a function of
F($\lambda$8446) in Figure \ref{figure2}. For 77\% of HAeBe stars, the emission strength of
\OI~$\lambda$8446 is stronger than \OI~$\lambda$7774.
Hence, recombination is not likely to be the dominant excitation mechanism for
the production of \OI~lines in HAeBe stars. 

\subsubsection{Collisional Excitation}

\begin{figure}
\figurenum{3}  
\includegraphics[width=8.5cm,height=8cm]{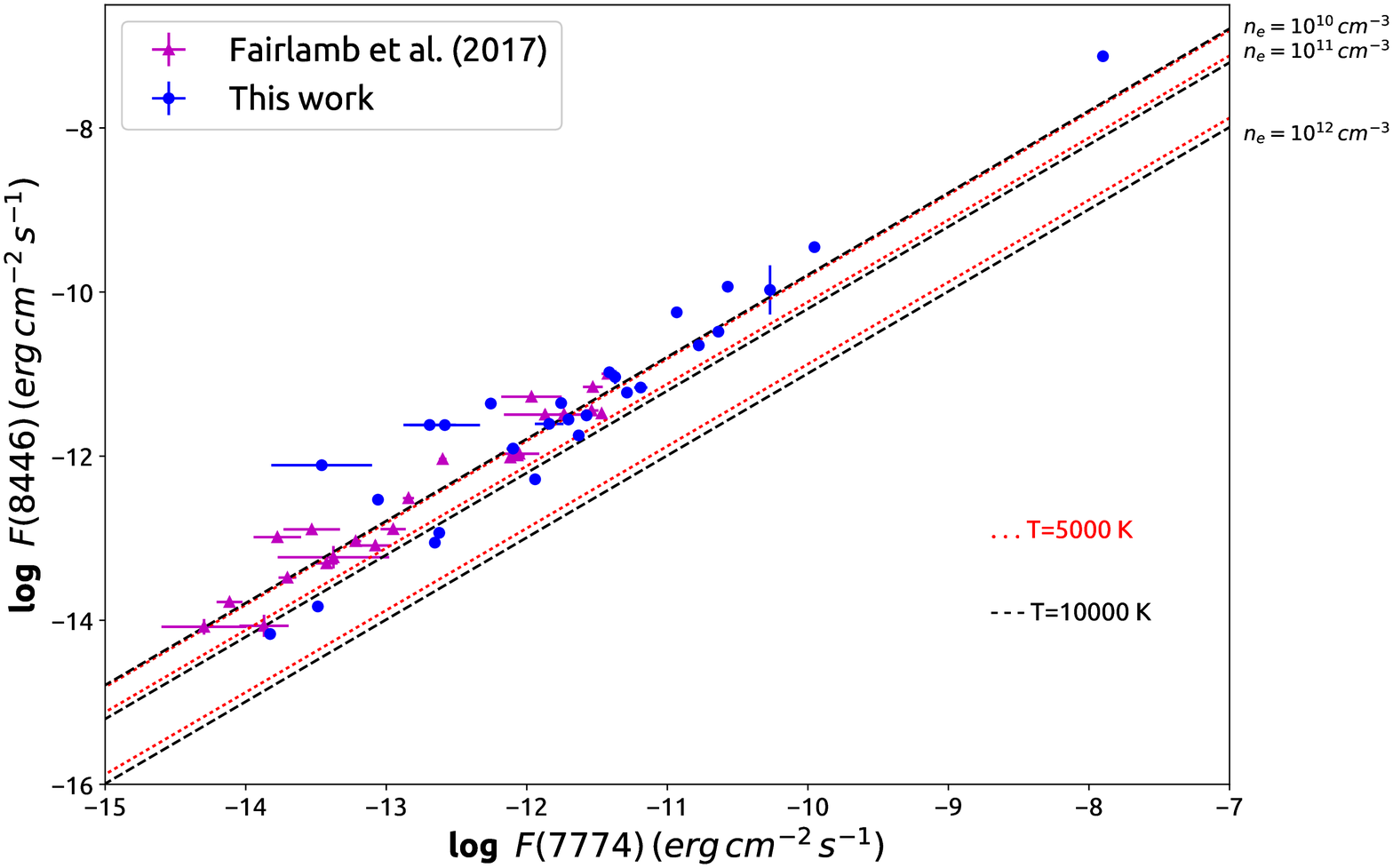}
\caption{Log-Log plot of F(8446) vs. F(7774): red dotted lines correspond to theoretical
  flux values for T = 5000 K and black
  dashed lines correspond to T = 10000 K, for n$_e$ = 10$^{10}$, 10$^{11}$, 10$^{12}$~cm$^{-3}$
  \citep{Kastner95}. The sources from \citet{Fairlamb17} are shown in purple triangles.}
\label{figure3}
\end{figure}

\citet{Bhatia95} built a hybrid model to compute the collisionally excited level populations and
line intensities of neutral oxygen under optically thin conditions. The intensities of all
possible allowed and forbidden \OI~lines in ultraviolet, visible and infrared wavelength regions
were calculated over a range of densities and temperatures seen in astrophysical systems.
\citet{Kastner95} estimated the expected values
of $\frac{F(8446)}{F(7774)}$ for various temperature, density combinations for collisional
excitation. In the magnetospheric accretion models for HAeBe
stars \citep[e.g.][]{Muzerolle04}, most of the emission lines observed in the visible
and near-IR wavelengths are formed in magnetospheric accretion columns. It is possible that
\OI~lines also form in these accretion columns. The typical accretion rates for HAeBe
stars are in the range of 1.0$\times$10$^{-8}$ -- 1.0$\times$10$^{-6}$ M$_\odot$ yr$^{-1}$
with a median value of $\sim$2.0$\times$10$^{-7}$ M$_\odot$ yr$^{-1}$
\citep[e.g.][]{Mendigutia11,Mendigutia12}. The corresponding density of
accretion columns are in the range of
10$^{11}$ -- 10$^{13}$~cm$^{-3}$ for temperatures of 6000 -- 10000 K,
for typical parameters of magnetospheric accretion
models \citep[see][]{Muzerolle04,Muzerolle98,Muzerolle01,Hartmann94}.
We have taken theoretical \OI~line flux ratio values corresponding
to these temperature, density combinations from \citet{Kastner95}.
Observational data is shown in Figure \ref{figure3} for
50 HAeBe stars, including the measurements of 30 sources
from \citet{Fairlamb17}. The flux values of \OI~$\lambda$7774 and $\lambda$8446
corresponding to a temperature of 5000 K and
densities of 10$^{10}$, $10^{11}$, $10^{12}$ cm$^{-3}$ are represented as dotted lines
in Figure \ref{figure3} and $T$ = 10,000~K, n$_e$ = 10$^{10}$, 10$^{11}$,
10$^{12}$~cm$^{-3}$ combinations are
shown in dashed lines. Figure \ref{figure3} shows that the observed flux ratio for
almost all the sources in our sample is greater than those
predicted for densities $>$ 10$^{11}$~cm$^{-3}$. Additionally,
models for H$\alpha$ emission in HBe stars also require densities
of n$_e$ = 2$\times$10$^{12}$ cm$^{-3}$ for the line forming region
\citep[see][]{Patel16,Patel17}. Although these studies do
not discuss about \OI~line forming region, the strong correlation
between H$\alpha$ and \OI~line emission (see Section 4.1.4)
indicates that both lines are formed in the same region. Thus, our analysis suggest
that collisional excitation may not be the prominent mechanism
at densities $>$ 10$^{11}$~cm$^{-3}$ seen in the line forming regions of
HAeBe stars.

We have included \OI~$\lambda$ $\lambda$7774, 8446 line measurements of a sample of HAeBe 
stars studied in \citet{Fairlamb17}. These objects were observed with 
X-shooter spectrograph mounted at Very Large Telescope, Chile. Figure \ref{figure3} shows that 
the inclusion of the sample of HAeBe stars from X-Shooter provides more data in the lower flux
regime of 7774 and 8446 lines. Further, \OI~$\lambda$8446 flux values are more intense than
the theoretical estimates corresponding to $T$ = 5000/10000 K 
and n$_e$ = $10^{11}$ cm$^{-3}$. This analysis strengthens the claim that collisional excitation is not the
dominant excitation mechanism for the production of \OI~emission lines in HAeBe stars. 

\begin{figure}
\figurenum{4}  
\includegraphics[width=8cm,height=7.5cm]{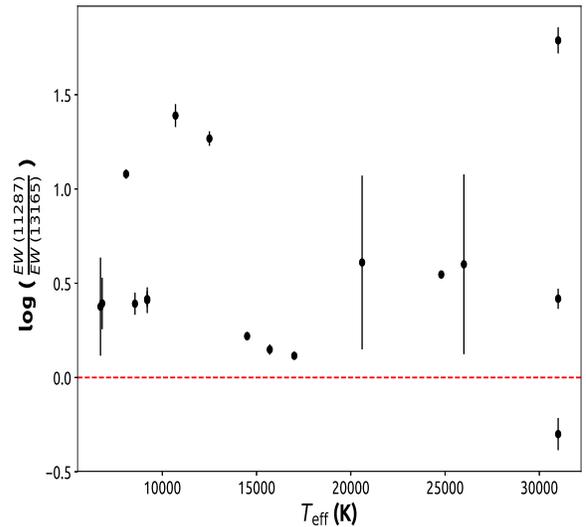}
\caption{Plot of ratio of EWs of \OI~$\lambda$11287 to \OI~$\lambda$13165
  against effective temperature of the star. It can be
  seen that in most cases $\frac{EW({11287})}{EW({13165})}>$1.  }
\label{figure4}
\end{figure}

Further confirmation is obtained from the analysis of the infrared spectra of HAeBe stars.
It has been proposed that if collisional excitation is the
dominant excitation mechanism, the equivalent width of $\lambda$13165 should be greater than that
of $\lambda$11287 i.e., $W(13165)/W(11287) \geq 1$ at T = 10,000 and 20,000 K,
respectively, for n$_e$ = 10$^{10}$ -- 10$^{12}$~cm$^{-3}$ \citep{Bhatia95}. For most of our sample of stars
we found that the emission line strength of \OI~$\lambda$11287 higher than that of
$\lambda$13165 (see Figure \ref{figure4}), confirming that collisional excitation does not play a major role
in the formation of \OI~emission lines in HAeBe stars. 

\subsubsection{Continuum fluorescence}

Continuum fluorescence was invoked as the excitation mechanism
for the production of \OI~lines in the spectra of planetary nebulae \citep{Seaton68} and Orion
Nebula \citep{Grandi75b}. Grandi in his thesis \citep{Grandi75a} and in a paper
summarizing the thesis results \citep{Grandi75b} showed that the expected theoretical
ratio of the line strengths of the $\lambda$13165 and $\lambda$11287 lines due to starlight excitation
(equivalently continuum fluorescence) should be of the order of 10
or slightly more. These model predictions are summarized in Tables 7 and Table 2
of \citet{Grandi75a} and \citet{Grandi75b} respectively and also
described in the text. In essence, $\lambda$13165 is predicted to be much stronger
than $\lambda$11287 if continuum 
fluorescence is the dominant excitation mechanism for the \OI~lines \citep[also see][]{Strittmatter77}.
This prediction by Grandi was confirmed observationally for the Orion nebula in the
spectroscopic studies by \citet{Lowe77}. Also, strong \OI~emission lines at 7002 \AA,
7254 \AA~and 7990 \AA~lines would be observed in the spectra
\citep{Strittmatter77,Grandi80}. Apart from the Orion nebula, another instance
where the $\lambda$13165 line is stronger than the $\lambda$11287 line
is  in the  inner 10 arcsecond sized nebula surrounding P Cygni. Near-infrared 1$-$2.5 micron spectra
by \citet{Smith06} of this region
gives a value  of 2.55$\pm$0.57 for the ratio of the 13165 \& 11287 line strengths.

Our analysis show that the emission
strength of \OI~$\lambda$11287 is greater than that of \OI~$\lambda$13165 for our
sample of HAeBe stars (Figure \ref{figure4}). In addition, we do not see emission lines
at 7002 \AA, 7254 \AA~and 7990 \AA~in any of the object spectra.
This suggests that continuum fluorescence is unlikely to be the dominant mechanism for the formation
of \OI~emission lines in HAeBe stars.

\subsubsection{Lyman $\beta$ fluorescence}

\begin{figure}
\figurenum{5}  
\includegraphics[width=7cm,height=7cm]{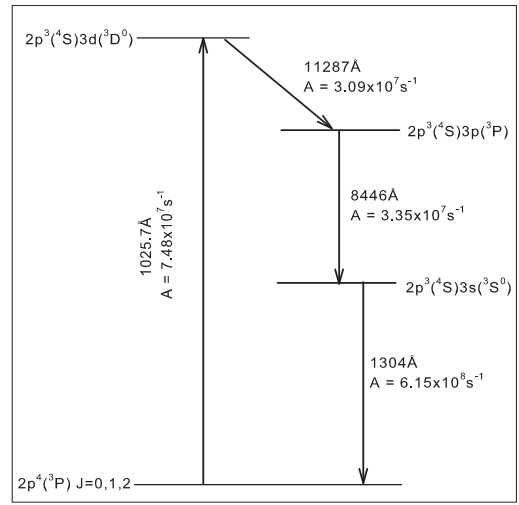}
\caption{Figure shows the pumping and the fluorescent transitions in \OI~caused
by the Lyman beta fluorescence process. }
\label{figure5}
\end{figure}

Lyman $\beta$ (Ly$\beta$) fluorescence occurs because the $3d^3D^0$ level of \OI~is populated
by Ly$\beta$ radiation, with subsequent cascades producing the \OI~$\lambda\lambda$11287,
8446, 1304 lines in emission (Figure \ref{figure5}). This is due to the near coincidence in wavelength of
Lyman beta and the \OI~resonance line 2$p~^3P_2$ -- 3$d~^3D^0_{321}$ at 1025.77 \AA~\citep{Bowen47}.  
Our analysis show that the cascade lines expected from Ly$\beta$ fluorescence,
\OI~$\lambda$8446 and \OI~$\lambda$11287 are quite strong in the spectra of HAeBe stars.
Also, \OI~$\lambda$7774 is less intense than \OI~$\lambda$8446, suggesting
that collisional excitation and recombination are relatively less important for \OI~excitation in HAeBe stars.
Similarly, the lower emission strength of \OI~$\lambda$13165 with
respect to $\lambda$11287 rules out collisional excitation and continuum fluorescence as the dominant mechanisms
for the production of \OI~lines. Further, \OI~$\lambda \lambda$7002, 7254,
7990 emission lines are not present in the spectra of
HAeBe stars. These lines are generally seen in sources where \OI~lines are excited by continuum fluorescence.
All these pieces of evidence strongly suggest that Ly$\beta$ fluorescence is likely to be
the dominant excitation mechanism for the production of \OI~lines in HAeBe stars.

\begin{figure}
\figurenum{6}
\includegraphics[width=8cm,height=8cm]{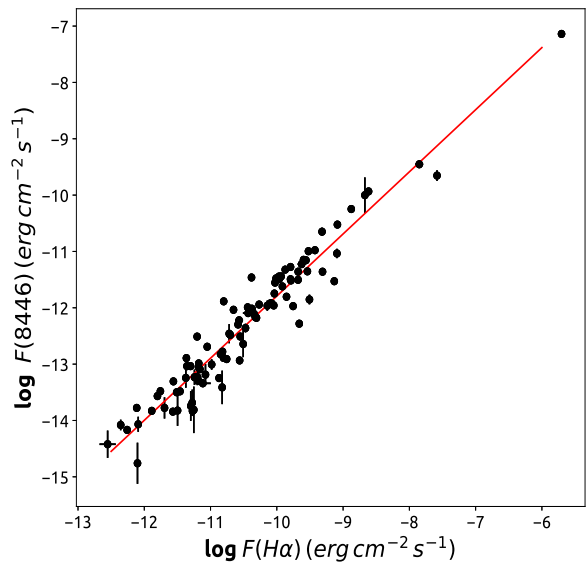}
\caption{Log-Log plot of F(8446) vs F(H$\alpha$): The sample of HAeBe stars from
  our studies and from \citet{Fairlamb17} are shown in black circles. We plotted
    97 sources including 57 from \citet{Fairlamb17}, of which 7 are common
    between \citet{Fairlamb17} and our sample.}
\label{figure6}
\end{figure}

If Ly$\beta$ fluorescence is responsible for \OI~emission in HAeBe stars, then one would
expect a correlation between H$\alpha$ and
\OI~$\lambda$8446 line intensities. Ly$\beta$ photon results from the n = 3$-$1
transition of the hydrogen atom and the H$\alpha$ photon results from the
transition n = 3$-$2. Thus the upper level of both the transitions are the same.
In other words, hydrogen atoms in the excited state of n = 3 are responsible for both
lines. If these lines originate from the same gas component, one would expect their
intensities to be correlated. If, in addition, \OI~$\lambda$8446 intensity is
proportional to Ly$\beta$ intensity, then one would expect a correlation between
H$\alpha$ and \OI~$\lambda$8446. This is shown in Figure \ref{figure6}, where F($\lambda$8446) is shown
as a function of F(H$\alpha$) for our sample and those from \citet{Fairlamb17}. A correlation in seen between the
flux values of H$\alpha$ and \OI~$\lambda$8446, suggesting the application of Ly$\beta$ fluorescence
process. A linear fit to the distribution of points in Figure \ref{figure6} gives a relation of the form
F($\lambda$8446)$\propto$F(H$\alpha$)$^{1.02\pm0.04}$, with a  Pearson's
correlation coefficient of 0.96. From the analysis it is clear that H$\alpha$ emission is correlated
with the emission strength of \OI~$\lambda$8446. This particularly has implications in understanding the region
of formation of \OI~emission lines in HAeBe stars. \\

\figurenum{6}  

If Ly$\beta$ photons and \OI~atoms do not co-exist, Ly$\beta$ fluorescence would not have
been possible since Ly$\beta$ gets scattered off neutral hydrogen resulting in
the production of Ly$\alpha$ and H$\alpha$, before interacting with neutral oxygen atoms. This is the reason why
Ly$\beta$ fluorescence does not operate in Orion nebula where Ly$\beta$ photons are trapped in
the inner regions of the nebula whereas the \OI~is confined in the exterior \citep{Grandi75b}.
In the case of HAeBe stars, H$\alpha$ is thought to originate in the magnetospheric
accretion columns which connects the inner disk to the central star. The fact that Ly$\beta$ fluorescence
operates in HAeBe stars suggests that \OI~emission lines are also formed in these
accretion columns in HAeBe stars.

It is worth noting that the gas has to be optically
thick in H$\alpha$ in order for \OI~$\lambda$8446 to get excited by Ly$\beta$ fluorescence.
From the equations of level populations in statistical equilibrium,
\citet{Grandi80} derived an optical depth in H$\alpha$ ($\tau_{H\alpha}$) of 2000,
considering Ly$\beta$ fluorescence operating in active galaxies. We examined whether the line forming region
in HAeBe stars are optically thick in H$\alpha$. For the sample of HAeBe stars considered in this study,
the median flux ratio F(H$\alpha$/$\lambda$8446)$_{obs}$ is 65.2. From the analysis of level populations,
\citet{Strittmatter77} derived a theoretical H$\alpha$ to \OI~$\lambda$8446 flux ratio
of around 7500, under optically thin conditions. The optical depth of H$\alpha$
can be estimated from the ratio of
theoretical to observed line flux ratio,
i.e., $\tau_{H\alpha}$ = 7500 / F(H$\alpha$/$\lambda$8446)$_{obs}$ = 115.
The sufficiently high value of optical depth in H$\alpha$ derived in the case of HAeBe stars agrees
with the assumption that the gas needs to be optically thick thereby trapping the Lyman beta photons,
paving way for Ly$\beta$ fluorescence. 

\section{Conclusion}

From an analysis of the observed optical spectra of 62 HAeBe stars
and near-infrared spectra of 17 HAeBe stars, we have shown that Ly$\beta$ fluorescence
is likely to be the dominant excitation mechanism for the formation of
\OI~emission lines. We ruled out recombination and continuum fluorescence as the
possible excitation mechanisms as the emission strength of \OI~$\lambda$8446
and $\lambda$11287 are much stronger than the adjacent \OI~lines at $\lambda$7774
and $\lambda$13165, respectively. We found that collisional excitation does not contribute substantially
to \OI~emission from the comparative analysis of the observed line flux values of $\lambda$7774
and $\lambda$8446 with those predicted by the theoretical models of \citet{Kastner95}.

\section*{Acknowledgments}

We would like to thank the referee for their comments which helped in
improving the quality of the manuscript. We would like
to thank the staff at IAO, Hanle and its remote
control station at CREST, Hosakote for their help during the observation
runs. This research uses the SIMBAD astronomical data base service operated at
CDS, Strasbourg. This publication made use data of 2MASS, which
is a joint project of University of Massachusetts and the Infrared
Processing and Analysis Centre/California Institute of Technology,
funded by the National Aeronautics and Space Administration and
the National Science Foundation.


\begin{thebibliography}{}
  
\bibitem[Ashok et al.(2006)]{Ashok06} Ashok, N. M., Banerjee, D. P. K., Varricatt, W. P., Kamath, U. S. 2006, MNRAS, 368, 592
\bibitem[Banerjee \& Ashok(2012)]{Banerjee12} Banerjee, D. P. K., Ashok, N. M. 2012, BASI, 40, 243
\bibitem[Bessell(1983)]{Bessell83} Bessell, M. S. 1983, PASP, 95, 480
\bibitem[Bhatia \& Kastner(1995)]{Bhatia95} Bhatia, A. K., Kastner, S. O. 1995, ApJS, 96, 325
\bibitem[Bowen(1947)]{Bowen47} Bowen, I. S. 1947, PASP, 59, 196
\bibitem[Briot(1981)]{Briot81} Briot, D. 1981, A\&A, 103, 5	
\bibitem[Cohen \& Kuhi(1979)]{Cohen79} Cohen, M., Kuhi, L. V. 1979, ApJS, 41, 743
\bibitem[Calvet \& Gullbring(1998)]{Calvet98} Calvet, N., Gullbring, E. 1998, ApJ, 509, 802
\bibitem[Carmona et al.(2010)]{Carmona10} Carmona, A., van den Ancker, M.~E., Audard, M., et al.\ 2010, A\&A, 517, A67 
\bibitem[Castelli et al.(1997)]{Castelli97} Castelli, F., Gratton, R. G., Kurucz, R. L. 1997, A\&A, 318, 841
\bibitem[Cauley \& Johns-Krull(2015)]{Cauley15} Cauley, P. W. \& Johns-Krull, C. M. 2015, ApJ, 810, 5
\bibitem[Corcoran \& Ray(1998)]{Corcoran98} Corcoran, M., Ray, T. P. 1998, A\&A, 331, 147
\bibitem[Dahm(2008)]{Dahm08} Dahm, S. E. 2008, AJ, 136, 521
\bibitem[Drew et al.(1997)]{Drew97} Drew, J.~E., Busfield, G., Hoare, M.~G., et al.\ 1997,MNRAS, 286, 538 
\bibitem[Fairlamb et al.(2017)]{Fairlamb17} Fairlamb, J. R., Oudmaijer, R. D., Mendigutia, I. et al. 2017, MNRAS, 464, 4721
\bibitem[Fairlamb et al.(2015)]{Fairlamb15} Fairlamb, J. R., Oudmaijer, R. D., Mendigutia, I. et al. 2015, MNRAS, 453, 976
\bibitem[Fang et al.(2009)]{Fang09} Fang, M., van Boekel, R., Wang, W. et al. 2009, A\&A, 504, 461
\bibitem[Felenbok et al.(1988)]{Felenbok88} Felenbok, P., Czarny, J., Catala, C., Praderie, F. 1988, A\&A, 201, 247
\bibitem[Ferland \& Netzer(1979)]{Ferland79} Ferland, G., Netzer, H. 1979, ApJ, 229, 274
\bibitem[Finkenzeller \& Mundt(1984)]{Finkenzeller84} Finkenzeller, U., Mundt, R. 1984, A\&AS, 55, 109
\bibitem[Gandolfi et al.(2008)]{Gandolfi08} Gandolfi, D., Alcal{\'a}, J.~M., Leccia, S. et al.\ 2008, ApJ, 687, 1303 
\bibitem[Garrison(1970)]{Garrison70} Garrison, R.~F.\ 1970, AJ, 75, 1001 
\bibitem[Gorti \& Bhatt(1993)]{Gorti93}	Gorti, U., Bhatt, H. C. 1993, A\&A, 270, 426
\bibitem[Grandi(1975a)]{Grandi75a} Grandi, S. A. 1975a, Ph.D. thesis, Univ. Arizona
\bibitem[Grandi(1975b)]{Grandi75b} Grandi, S. A. 1975b, ApJ, 196, 465
\bibitem[Grandi(1980)]{Grandi80} Grandi, S. A. 1980, ApJ, 238, 10
\bibitem[Gullbring et al.(1998)]{Gullbring98} Gullbring, E., Hartmann, L., Briceno, C., Calvet, N. 1998, ApJ, 492, 323
\bibitem[Hamann \& Persson(1992)]{Hamann92} Hamann, F., Persson, S. E. 1992, ApJS, 82, 247
\bibitem[Hartmann et al.(1994)]{Hartmann94} Hartmann, L., Hewett, R., Calvet, N. 1994, ApJ, 426, 669
\bibitem[Hauschildt et al.(1999)]{Hauschildt99} Hauschildt, P. H., Allard, F., Baron, E. 1999, ApJ, 512, 377
\bibitem[Herbig(1960)]{Herbig60} Herbig G. H., 1960, ApJS, 4, 337 
\bibitem[Herczeg \& Hillenbrand(2008)]{Herczeg08} Herczeg, G. J. \& Hillenbrand, L. A. 2008, ApJ, 681, 594
\bibitem[Hernandez et al.(2004)]{Hernandez04} Hern{\'a}ndez, J., Calvet, N., Brice{\~n}o, C. et al. 2004, AJ, 127, 1682
\bibitem[Hillenbrand et al.(1992)]{Hillenbrand92} Hillenbrand, L. A., Strom, S. E., Vrba, F. J., Keene, J. 1992, ApJ, 397, 613
\bibitem[Ingleby et al.(2013)]{Ingleby13} Ingleby, L., Calvet, N., Herczeg, G. et al. 2013, ApJ, 767, 112 
\bibitem[Kastner \& Bhatia(1995)]{Kastner95} Kastner, S. O., \& Bhatia, A. K. 1995, ApJ, 439, 346
\bibitem[Kenyon \& Hartmann(1995)]{Kenyon95} Kenyon, S. J., Hartmann, L. 1995, ApJS, 101, 117
\bibitem[Kurucz(1993)]{Kurucz93} Kurucz R. 1993, ATLAS9 Stellar Atmosphere Programs and 2 km/s grid, Kurucz CD-ROM No. 13. Smithsonian Astrophysical Observatory, Cambridge, MA
\bibitem[Lee \& Chen(2007)]{Lee07} Lee, H.-T., \& Chen, W.~P.\ 2007, ApJ, 657, 884 
\bibitem[Liu et al.(2011)]{Liu11} Liu, T., Zhang, H. Wu, Y. et al. 2011, ApJ, 734, 22
\bibitem[Lowe, Moorhead \& WehlauLowe (1977)]{Lowe77} Lowe, R. P., Moorhead, J. M., Wehlau, W. H. 1977, ApJ, 214, 712
\bibitem[Manoj et al.(2006)]{Manoj06} Manoj, P., Bhatt, H. C., Maheswar, G., Muneer, S. 2006, ApJ, 653, 657
\bibitem[Mathew et al.(2012a)]{Mathew12a} Mathew, B., Banerjee, D. P. K., Naik, S., Ashok, N. M., 2012a, MNRAS, 423, 2486
\bibitem[Mathew et al.(2012b)]{Mathew12b} Mathew, B., Banerjee, D. P. K., Subramaniam, A., Ashok, N. M., 2012b, ApJ, 753, 13
\bibitem[McDonald et al.(2017)]{McDonald17} McDonald, I., Zijlstra, A.~A., \& Watson, R.~A.\ 2017, VizieR Online Data Catalog, 747,
\bibitem[Mora et al.(2001)]{Mora01} Mora, A., Mer{\'{\i}}n, B., Solano, E., et al.\ 2001, A\&A, 378, 116 
\bibitem[Mathis(1990)]{Mathis90} Mathis, J. S. 1990, ARA\&A, 28, 37
\bibitem[Mendigut\'ia et al.(2011)]{Mendigutia11} Mendigut\'ia, I., Calvet, N., Montesinos, B. et al. 2011, A\&A, 535, A99
\bibitem[Mendigut\'ia et al.(2012)]{Mendigutia12} Mendigut\'ia I., Mora A., Montesinos B. et al. 2012, A\&A, 543, A59
\bibitem[McClure(2009)]{McClure09} McClure, M. 2009, ApJ, 693, L81
\bibitem[Munari et al.(2005)]{Munari05} Munari, U., Sordo, R., Castelli, F., \& Zwitter, T. 2005, A\&A, 442, 1127
\bibitem[Muzerolle et al.(2004)]{Muzerolle04} Muzerolle, J., D'Alessio, P., Calvet, N., \& Hartmann, L. 2004, ApJ, 617, 406
\bibitem[Muzerolle et al.(2001)]{Muzerolle01} Muzerolle, J., Calvet, N., Hartmann, L. 2001, ApJ, 550, 944
\bibitem[Muzerolle et al.(1998)]{Muzerolle98} Muzerolle, J., Calvet, N., Hartmann, L. 1998, ApJ, 492, 743
\bibitem[Patel et al.(2017)]{Patel17} Patel, P., Sigut, T. A. A., Landstreet, J. D. 2017, ApJ, 836, 214
\bibitem[Patel et al.(2016)]{Patel16} Patel, P., Sigut, T. A. A., Landstreet, J. D. 2016, ApJ, 817, 29
\bibitem[Pecaut \& Mamajek(2013)]{Pecaut13} Pecaut, M. J., Mamajek, E. E. 2013, ApJS, 208, 9
\bibitem[Porter \& Rivinius(2003)]{Porter03} Porter, J. M., \& Rivinius, T. 2003, PASP, 115, 1153
\bibitem[Seaton(1968)]{Seaton68} Seaton, M. J. 1968, MNRAS, 139, 129
\bibitem[Shu et al.(1994)]{Shu94} Shu, F., Najita, J., Ostriker, E. et al. 1994, ApJ, 429, 781
\bibitem[Slettebak(1951)]{Slettebak51} Slettebak, A. 1951, ApJ, 113, 436 
\bibitem[Smith \& Hartigan(2006)]{Smith06} Smith, N., Hartigan, P. 2006, ApJ, 638, 1045
\bibitem[Skiff(2014)]{Skiff14} Skiff, B.~A.\ 2014, VizieR Online Data Catalog, 1,  
\bibitem[Strittmatter et al.(1977)]{Strittmatter77} Strittmatter, P. A., Woolf, N. J., Thompson, R. I. et al. 1977, ApJ, 216, 23
\bibitem[The et al.(1994)]{The94} The, P. S., de Winter, D., Perez, M. R. 1994, A\&AS, 104, 315
\bibitem[Tjin A Djie et al.(2001)]{Tjin01} Tjin A Djie, H.~R.~E., van den Ancker, M.~E., Blondel, P.~F.~C. et al.\ 2001, MNRAS, 325, 1441
\bibitem[Vieira et al.(2003)]{Vieira03} Vieira, S. L. A., Corradi, W. J. B., Alencar, S. H. P. et al. 2003, AJ, 126, 2971
\bibitem[Waters \& Waelkens(1998)]{Waters98} Waters L. B. F. M., Waelkens, C., 1998, ARA\&A, 36, 233
\bibitem[Zhong et al.(2015)]{Zhong15} Zhong, J., L{\'e}pine, S., Li, J. et al.\ 2015, Research in Astronomy and Astrophysics, 15, 1154
\bibitem[Zacharias et al.(2004)]{Zacharias04} Zacharias, N., Monet, D.~G., Levine, S.~E. et al.\ 2004, Bulletin of the American Astronomical Society, 36, 48  

\end{thebibliography}
\end{document}